\documentclass[aoas,nameyear,seceqn,dvips]{arximspdf}
\usepackage{dcolumn,mathbh}
\usepackage{graphicx}

\doi{10.1214/10-AOAS338}
\volume{4}
\issue{3}
\pubyear{2010}
\firstpage{1158}
\lastpage{1182}

\makeatletter
\newproclaim{Rem}{Remark}
\newcolumntype{d}[1]{D{.}{.}{#1}}
\newcommand{\indicator}[1]{\mathbh{1}_{[{#1}] }}
\newcommand{\av}{\bolds{\alpha}}

\newcommand{\ev}{\mathbf{e}}

\newcommand{\fv}{\mathbf{f}}

\newcommand{\Gv}{\mathbf{G}}

\newcommand{\pv}{\mathbf{p}}
\newcommand{\hv}{\mathbf{h}}
\newcommand{\gv}{\mathbf{g}}

\newcommand{\qv}{\mathbf{q}}
\newcommand{\Mv}{\mathbf{M}}
\newcommand{\mv}{\bolds{\mu}}
\newcommand{\mvg}{\bolds{\mu}_g}

\newcommand{\yvg}{\mathbf{y}_g}
 \newcommand{\Zv}{\mathbf{Z}}

\newcommand{\muLS}{\hat{\mv}}
\newcommand{\SigmaLS}{\hat{\Sigma}}
\newcommand{\fvPanel}{\fv^{\,\mathrm{panel}}}
\newcommand{\SigmaPanel}{\Sigma^{\,\mathrm{panel}}}
\makeatother

\begin{document}
\begin{frontmatter}

\title{Using linear predictors to impute allele frequencies from
summary or pooled\break genotype data}
\runtitle{Imputing Allele Frequencies using Linear Predictors}

\begin{aug}
\author[A]{\fnms{Xiaoquan} \snm{Wen}\ead[label=e1]{wen@uchicago.edu}}
\and
\author[B]{\fnms{Matthew} \snm{Stephens}\thanksref{T1}\corref{}\ead[label=e2]{mstephens@uchicago.edu}}
\thankstext{T1}{Supported in part by NIH Grants HG02585 and HL084689.}
\runauthor{X. Wen and M. Stephens}
\affiliation{University of Chicago}
\address[A]{Department of Statistics\\
University of Chicago\\
Chicago, Illinois 60637\\
USA\\
\printead{e1}}
\address[B]{Department of Statistics\\
and \\
Department of Human Genetics\\
University of Chicago\\
Chicago, Illinois 60637\\
USA\\
 \printead{e2}}
\end{aug}

\received{\smonth{8} \syear{2009}}
\revised{\smonth{1} \syear{2010}}

\begin{abstract}
Recently-developed genotype imputation methods are a powerful tool for
detecting untyped genetic variants that affect disease susceptibility
in genetic association studies. However, existing imputation methods
require individual-level genotype data, whereas, in practice, it is
often the case that only summary data are available. For example, this
may occur because, for reasons of privacy or politics, only summary
data are made available to the research community at large; or because
only summary data are collected, as in DNA pooling experiments. In this
article we introduce a new statistical method that can accurately infer
the frequencies of untyped genetic variants in these settings, and
indeed substantially improve frequency estimates at typed variants in
pooling experiments where observations are noisy. Our approach, which
predicts each allele frequency using a linear combination of observed
frequencies, is statistically straightforward, and related to a long
history of the use of linear methods for estimating missing values
(e.g., Kriging). The main statistical novelty is our approach to
regularizing the covariance matrix estimates, and the resulting linear
predictors, which is based on methods from population genetics. We find
that, besides being both fast and flexible---allowing new problems to
be tackled that cannot be handled by existing imputation approaches
purpose-built for the genetic context---these linear methods are also
very accurate. Indeed, imputation accuracy using this approach is
similar to that obtained by state-of-the-art imputation methods that
use individual-level data, but at a fraction of the computational cost.
\end{abstract}


\end{frontmatter}
\section{Introduction}

Genotype imputation [\citet{servinstephens}, \citet{guanstephens}, \citet{marchinietal},
\citet{howieetal}, \citet{browning}, \citet{huangetal}] has recently emerged as a useful tool
in the analysis of genetic association studies as a way of performing
tests of association at genetic variants (specifically, Single
Nucleotide Polymorphisms, or SNPs) that were not actually measured in
the association study. In brief, the idea is to exploit the fact that
untyped SNPs are often correlated, in a known way, with one or more
typed SNPs. Imputation-based approaches exploit these correlations,
using observed genotypes at typed SNPs to estimate, or impute,
genotypes at untyped SNPs, and then test for association between the
imputed genotypes and phenotype, taking account of uncertainty in the
imputed genotypes. (Although in general statistics applications the
term ``imputation'' may imply replacing unobserved data with a single
point estimate, in the genetic context it is often used more broadly to
include methods that consider the full conditional distribution of the
unobserved genotypes, and this is the way we use it here.) These
approaches have been shown to increase overall power to detect
associations by expanding the number of genetic variants that can be
tested for association [\citet{servinstephens}, \citet{marchinietal}], but
perhaps their most important use has been in performing meta-analysis
of multiple studies that have typed different, but correlated, sets of
SNPs [e.g., \citet{zegginietal}].

Existing approaches to imputation in this context have been developed
to work with individual-level data: given genotype data at typed SNPs
in each individual, they attempt to impute the genotypes of each
individual at untyped SNPs. From a general statistical viewpoint, one
has a large number of correlated discrete-valued random variables
(genotypes), whose means and covariances can be estimated, and the aim
is to predict values of a subset of these variables, given observed
values of all the other variables. Although one could imagine applying
off-the-shelf statistical methods to this problem [e.g., \citet{yuschaid} consider approaches based on linear regression], in
practice, the most successful methods in this context have used
purpose-built methods based on discrete Hidden Markov Models (HMMs)
that capture ideas from population genetics [e.g., \citet{listephens},
\citet{scheetstephens}].

In this paper we consider a related, but different, problem: given the
\textit{frequency} of each allele at all typed SNPs, we attempt to
impute the \textit{frequency} of each allele at each untyped SNP. We
have two main motivations for considering this problem. The first is
that, although most large-scale association studies collect
individual-level data, it is often the case that, for reasons of
privacy [\citet{homeretalmixture}, \citet{sankararamanetal}] or politics, only
the allele frequency data (e.g., in cases vs. controls) are made
available to the research community at large. The second motivation is
an experimental design known as DNA pooling [\citet{homeretalpooling},
\citet{meaburnetal}], in which individual DNA samples are grouped into
``pools'' and high-throughput genotypings are performed on each pool.
This experimental design can be considerably cheaper than separately
genotyping each individual, but comes at the cost of providing only
(noisy) estimates of the allele frequencies in each pool. In this
setting the methods described here can provide not only estimates of
the allele frequencies at untyped SNPs, but also more accurate
estimates of the allele frequencies at typed SNPs.

From a general statistical viewpoint, this problem of imputing
frequencies is not so different from imputing individual genotypes:
essentially, it simply involves moving from discrete-valued variables
to continuous ones. However, this change to continuous variables
precludes direct use of the discrete HMM-based methods that have been
applied so successfully to impute individual genotypes. The methods we
describe here come from our attempts to extend and modify these
HMM-based approaches to deal with continuous data. In doing so, we end
up with a considerably simplified method that might be considered an
off-the-shelf statistical approach: in essence, we model the allele
frequencies using a multivariate normal distribution, which results in
unobserved frequencies being imputed using linear combinations of the
observed frequencies (as in Kriging, for example). Some connection with
the HMM-based approaches remains though, in how we estimate the mean
and variance--covariance matrix of the allele frequencies. In
particular, consideration of the HMM-based approaches lead to a natural
way to regularize the estimated variance--covariance matrix, making it
sparse and banded: something that is important here for both
computational and statistical reasons. The resulting methods are highly
computationally efficient, and can easily handle very large panels
(phased or unphased). They are also surprisingly accurate, giving
estimated allele frequencies that are similar in accuracy to those
obtained from state-of-the-art HMM-based methods applied to individual
genotype data. That is, one can estimate allele frequencies at untyped
SNPs almost as accurately using only the \textit{frequency} data at
typed SNPs as using the \textit{individual} data at typed SNPs.
Furthermore, when individual-level data are available, one can also
apply our method to imputation of individual genotypes (effectively by
treating each individual as a pool of 1), and this results in
imputation accuracy very similar to that of state-of-the-art HMM-based
methods, at a fraction of the computational cost. Finally, in the
context of noisy data from pooling experiments, we show via simulation
that the method can produce substantially more accurate estimated
allele frequencies at genotyped markers.

\section{Methods and models}

We begin by introducing some terminology and notation.

A \textit{SNP} (Single Nucleotide Polymorphism) is a genetic marker
that usually exhibits two different types (alleles) in a population. We
will use 0 and 1 to denote the two alleles at each SNP, with the
labeling being essentially arbitrary.

 A \textit{haplotype} is a
combination of alleles at multiple SNPs residing on a single copy of a
genome. Each haplotype can be represented by a string of binary (0/1)
values. Each individual has two haplotypes, one inherited from each
parent. Routine technologies read the two haplotypes simultaneously to
produce a measurement of the individual's genotype at each SNP, which
can be coded as 0, 1 or 2 copies of the 1 allele. Thus, the haplotypes
themselves are not usually directly observed, although they can be
inferred using statistical methods [\citet{stephenssmithdonnelly}].
Genotype data where the haplotypes are treated as unknown are referred
to as ``unphased,'' whereas if the haplotypes are measured or estimated,
they are referred to as ``phased.''

 In this paper we consider the following form of imputation problem. We assume that data are
available on $p$ SNPs in a reference panel of data on $m$ individuals
samples from a population, and that a subset of these SNPs are typed on
a further study sample of individuals taken from a similar population.
The goal is to estimate data at untyped SNPs in the study sample, using
the information on the correlations among typed and untyped SNPs that
is contained in the reference panel data.

We let $\mathbf{M}$ denote the panel data, and $\mathbf{h}_1,\ldots,\mathbf{h}_{2n}$ denote the $2n$ haplotypes in the study sample. In the
simplest case, the panel data will be a $2m \times p$ binary matrix,
and the haplotypes $\mathbf{h}_1,\ldots, \mathbf{h}_{2n}$ can be
thought of as additional rows of this matrix with some missing entries
whose values need ``imputing.'' Several papers have focused on this
problem of ``individual-level'' imputation [\citet{servinstephens},
\citet{scheetstephens}, \citet{marchinietal}, \citet{browning}, \citet{abecasis}].
 In this paper we consider the problem of
performing imputation when only summary-level data are available for
$\mathbf{h}_1,\ldots, \mathbf{h}_{2n}$. Specifically, let
\begin{equation}
  \label{jointf}
 \mathbf{y}=\pmatrix{y_1&\cdots&y_p}^\prime=\frac{1}{2n}\sum_{i=1}^{2n}\mathbf{h}_i
\end{equation}
denote the vector of allele frequencies in the study sample. We assume
that these allele frequencies are measured at a subset of
\textit{typed} SNPs, and consider the problem of using these
measurements, together with information in $\mathbf{M}$, to estimate
the allele frequencies at the remaining \textit{untyped} SNPs. More
formally, if $(\mathbf{y}_t,\mathbf{y}_u)$ denotes the partition of
$\mathbf{y}$ into typed and untyped SNPs, our aim is to estimate the
conditional distribution $ \mathbf{y}_u|\mathbf{y}_t, \mathbf{M}$.

Our approach is based on the assumption that $\mathbf{h}_1,\ldots,\mathbf{h}_{2n}$ are independent and identically distributed (i.i.d.)
draws from some conditional distribution
$\operatorname{Pr}(\mathbf{h}|\mathbf{M})$. Specifically, in common
with many existing approaches to individual-level imputation
[\citet{stephensscheet}, \citet{marchinietal}, \citet{abecasis}], we use the HMM-based
conditional distribution from \citet{listephens}, although other choices
could be considered. It then follows by the central limit theorem,
provided that the sample size $2n$ is large, the distribution of
$\mathbf{y}|\mathbf{M}$ can be approximated by a multivariate normal
distribution:
\begin{equation}
   \label{basicm}
   \mathbf{y}|\mathbf{M}\sim{\mathrm{N}}_p(\bolds{\mu},\Sigma),
\end{equation}
 where $\bolds{\mu}=\mathrm{E}(\mathbf{h}|\mathbf{M})$ and $\Sigma=\frac{1}{2n}\operatorname{Var}(\mathbf{h}|\mathbf{M})$.

 From this joint distribution, the required conditional
distribution is easily obtained. Specifically, by partitioning
$\bolds{\mu}$ and $\Sigma$ in the same way as $\mathbf{y}$, according
to SNPs' typed/untyped status, (\ref{basicm}) can be written
\begin{equation}
  \left ( \matrix{
    \mathbf{y}_u \cr
    \mathbf{y}_t } \bigg | \mathbf{M} \right) \sim \mathrm{N}_p
    \left(\pmatrix{\bolds{\mu}_u
   \cr\bolds{\mu}_t},
  \pmatrix{
    \Sigma_{uu} & \Sigma_{ut}
    \cr\Sigma_{tu} & \Sigma_{tt}}\right)
\end{equation}
 and
\begin{equation}
  \label{basicm_sol}
  \mathbf{y}_u|\mathbf{y}_t,\mathbf{M} \sim \mathrm{N}_q\bigl(\bolds{\mu}_u + \Sigma_{ut}\Sigma^{-1}_{tt}(\mathbf{y}_t-\bolds{\mu}_t), \Sigma_{uu} - \Sigma_{ut}\Sigma^{-1}_{tt}\Sigma_{tu}\bigr).
\end{equation}
The mean of this last distribution can be used as a point estimate for
the unobserved frequencies $\mathbf{y}_u$, while the variance gives an
indication of the uncertainty in these estimates. (In principle, the
mean can lie outside the range $[0,1)$, in which case we use 0 or 1, as
appropriate, as the point estimate; however, this happens very rarely
in practice.)

\begin{figure}

\includegraphics{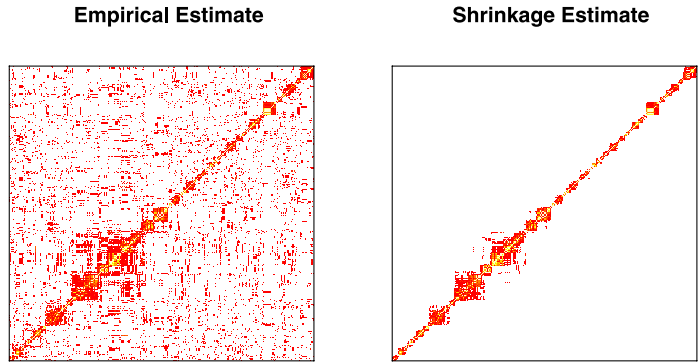}

\caption{Comparison of
empirical and shrinkage estimates (based on Li and Stephens Model) of
squared correlation matrix from the panel. Both of them are estimated
using Hapmap CEU panel with 120 haplotypes. The region plotted is on
chromosome 22 and contains 1000 Affymetrix SNPs which cover a 15 Mb
genomic region. Squared correlation values in $[0.05, 1.00]$ are
displayed using R's heat.colors scheme, with gold color representing
stronger correlation and red color representing weaker correlation.}\label{var.cmp.plot}
\end{figure}

The parameters $\bolds{\mu}$ and $\Sigma$ must be estimated from the
panel data. It may seem natural to estimate these using the empirical
mean $\mathbf{f}^{\,\mathrm{panel}}$ and the empirical covariance matrix
$\Sigma^{\mathrm{panel}}$ from the panel. However, $\Sigma^{\mathrm{panel}}$ is highly
rank deficient because the sample size $m$ in the panel is far less
than the number of SNPs $p$, and so this empirical matrix cannot be
used directly. Use of the conditional distribution from Li and Stephens
solves this problem. Indeed, under this conditional distribution
$\mathrm{E}(\mathbf{h}|\mathbf{M})=\hat{\bolds{\mu}}$ and $\operatorname{Var}(\mathbf{h}|\mathbf{M})=\hat{\Sigma}$ can be derived analytically
(Appendix \ref{ln.appx}) as
\begin{eqnarray}
  \label{basicm_mean}
  \hat{\bolds{\mu}}&=&(1-\theta)\mathbf{f}^{\,\mathrm{panel}} +\frac{\theta}{2}{\mathbf 1}, \\
  \label{basicm_var}
  \hat{\Sigma}&=&(1-\theta)^2 S + \frac{\theta}{2}\biggl(1-\frac{\theta}{2}\biggr) I,
\end{eqnarray}
 where $\theta$ is a parameter relating to mutation, and $S$ is
obtained from $\Sigma^{\mathrm{panel}}$ by shrinking off-diagonal entries toward
0. Specifically,
\begin{equation}
  \label{covm}
  S_{ij} =\cases{\Sigma^{\mathrm{panel}}_{ij},&\quad$i=j$,
  \cr\exp\biggl(-\displaystyle{\frac{\rho_{ij}}{2m}}\biggr)
\Sigma^{\mathrm{panel}}_{ij},&\quad$i \ne j,$}
   \end{equation}
      where $\rho_{ij}$ is an estimate of the population-scaled recombination
rate between SNPs $i$ and $j$ [e.g.,
\citet{hudson01}, \citet{listephens}, \citet{mcveanetal}]. We use the value of $\theta$
suggested by \citet{listephens},
\begin{equation}
  \theta = \frac{(\sum_{i=1}^{2m-1}{1}/{i})^{-1}}{2m + (\sum_{i=1}^{2m-1}{1}/{i})^{-1}},
\end{equation}
 and values of $\rho_{ij}$ obtained by applying the software
PHASE [\citet{stephensscheet}] to the HapMap CEU data, which are
conveniently distributed with the \mbox{IMPUTE} software package. For SNPs $i$
and $j$ that are distant, $\exp(-\frac{\rho_{ij}}{2m}) \approx 0 $. To
exploit the benefits of sparsity, we set any value that was less than
$10^{-8}$ to be 0, which makes $\hat{\Sigma}$  sparse and banded: see
Figure \ref{var.cmp.plot} for illustration. This makes matrix inversion
in (\ref{basicm_sol}) computationally feasible and fast, using standard
Gaussian elimination.

\subsection{Incorporating measurement error and overdispersion}

Our treatment above assumes that the allele frequencies of typed SNPs, $\mathbf{y}_t$, are observed without error. In some settings,
for example, in DNA pooling experiments, this is not the case. We incorporate measurement error by introducing a single parameter
$\varepsilon^2$, and assume
\begin{equation}
 \label{exp_error}
  \mathbf{y}_t^{\mathrm{obs}}|\mathbf{y}_t^{\mathrm{true}} \sim\mathrm{N}_{p-q}(\mathbf{y}_t^{\mathrm{true}}, \varepsilon^2 I ),
\end{equation}
where random vectors $\mathbf{y}_t^{\mathrm{obs}}$ and $\mathbf{y}_t^{\mathrm{true}}$ represent the observed and true sample allele frequencies for typed
SNPs respectively, and subscript $p-q$ denotes the number of typed SNPs. We assume that, given $\mathbf{y}_t^{\mathrm{true}}$, the observations
$\mathbf{y}_t^{\mathrm{obs}}$ are conditionally independent of the panel data ($\mathbf{M}$) and the allele frequencies at untyped SNPs ($\mathbf{y}_u^{\mathrm{true}}$).

Our treatment in the previous section also makes several other implicit assumptions: for example, that the panel and study individuals are
sampled from the same population, and that the parameters $\rho$ and $\theta$ are estimated without error. Deviations from these assumptions
will cause overdispersion: the true allele frequencies will lie further from their expected values than the model predicts. To allow for
this, we modify (\ref{basicm}) by introducing an overdispersion parameter $\sigma^2$:
\begin{equation}
  \label{sys_error}
  \mathbf{y}^{\mathrm{true}}|\mathbf{M}\sim\mathrm{N}_p(\hat{\bolds{\mu}}, \sigma^2\hat{\Sigma}).
\end{equation}
Overdispersion models like this are widely used for modeling binomial data [\citet{mccullaghnelder}].

In our applications below, for settings involving measurement error (i.e., DNA pooling experiments), we estimate $\sigma^2$, $\varepsilon^2$
by maximizing the multivariate normal likelihood:
\begin{equation}
 \label{obs.likelihood}
  \mathbf{y}_t^{\mathrm{obs}}|\mathbf{M} \sim \mathrm{N}_{p-q}(\hat{\bolds{\mu}}_t,\sigma^2 \hat{\Sigma}_{tt} + \varepsilon^2 I ).
\end{equation}
 For settings without measurement error, we set $\varepsilon^2=0$ and estimate $\sigma^2$ by maximum
 likelihood.

From the hierarchical model defined by (\ref{exp_error}) and (\ref{sys_error}), the conditional distributions of allele
frequencies at untyped and typed SNPs are given by
\begin{eqnarray}
  \label{improvm}
  \mathbf{y}_u^{\mathrm{true}}|\mathbf{y}_t^{\mathrm{obs}}, \mathbf{M}&\sim&{\mathrm{N}}_q\biggl(\hat{\bolds{\mu}}_u +
  \hat{\Sigma}_{ut}\biggl(\hat{\Sigma}_{tt}+\frac{\varepsilon^2}{\sigma^2}I\biggr)^{-1}(\mathbf{y}_t^{\mathrm{obs}}-\hat{\bolds{\mu}}_t),\nonumber\\[-8pt]\\[-8pt]
&&\hspace*{26pt}\sigma^2 \biggl(\hat{\Sigma}_{uu} -
\hat{\Sigma}_{ut}\biggl(\hat{\Sigma}_{tt}+\frac{\varepsilon^2}{\sigma^2}I\biggr)^{-1}\hat{\Sigma}_{tu}\biggr)\biggr),\nonumber
\end{eqnarray}
and
\begin{eqnarray}
  \label{nrd}
   \qquad\mathbf{y}_t^{\mathrm{true}}|\mathbf{y}_t^{\mathrm{obs}},\mathbf{M} &\sim&\mathrm{N}_{p-q}\biggl(\biggl(\frac{1}{\sigma^2}\hat{\Sigma}_{tt}^{-1}+\frac{1}{\varepsilon^2}I\biggr)^{-1}\biggl(\frac{1}{\sigma^2}\hat{\Sigma}_{tt}^{-1} \hat{\bolds{\mu}}_t + \frac{1}{\varepsilon^2}
   \mathbf{y}_t^{\mathrm{obs}}\biggr),\nonumber\\[-8pt]\\[-8pt]
                                                               && \hspace*{136pt}\biggl(\frac{1}{\sigma^2}\hat{\Sigma}_{tt}^{-1}+\frac{1}{\varepsilon^2}I\biggr)^{-1}\biggr).\nonumber
\end{eqnarray}
We use (\ref{improvm}) to impute allele frequencies at untyped SNPs. In particular, we use the conditional mean
\begin{equation}
\label{point.est.untyped}
  \hat\mathbf{y}^{\mathrm{true}}_u = \hat{\bolds{\mu}}_u + \hat{\Sigma}_{ut}\biggl(\hat{\Sigma}_{tt}+\frac{\varepsilon^2}{\sigma^2}I\biggr)^{-1}(\mathbf{y}_t^{\mathrm{obs}}-\hat{\bolds{\mu}}_t)
\end{equation}
as a natural point estimate for these allele frequencies. In settings involving measurement error, we use (\ref{nrd}) to estimate
allele frequencies at typed SNPs, again using the mean
\begin{equation}
\label{point.est.typed}
  \hat \mathbf{y}^{\mathrm{true}}_t = \biggl(\frac{1}{\sigma^2}\hat{\Sigma}_{tt}^{-1}+\frac{1}{\varepsilon^2}I\biggr)^{-1}\biggl(\frac{1}{\sigma^2}\hat{\Sigma}_{tt}^{-1} \hat{\bolds{\mu}}_t + \frac{1}{\varepsilon^2} \mathbf{y}_t^{\mathrm{obs}}\biggr)
\end{equation}
 as a point estimate. Note that this mean has an intuitive interpretation as a weighted average of the observed allele frequency at
that SNP and information from other nearby, correlated, SNPs. For example, if two SNPs are perfectly correlated, then in the presence
of measurement error, the average of the measured frequencies will be a better estimator of the true frequency than either of the single
measurements (assuming measurement errors are uncorrelated). The lower the measurement error, $\varepsilon^2$, the greater the weight given
to the observed frequencies; and when $\varepsilon^2=0$ the estimated frequencies are just the observed frequencies.

\begin{Rem*}For both untyped and typed SNPs, our point estimates for allele frequencies, (\ref{point.est.untyped}) and
(\ref{point.est.typed}), are linear functions of the observed allele frequencies. Although these linear predictors were
developed based on an appeal to the Central Limit Theorem, and resultant normality assumption, there are alternative
justifications for use of these particular linear functions that do not rely on normality. Specifically, assuming that the two
haplotypes making up each individual are i.i.d. draws from a conditional distribution $\operatorname{Pr}(\mathbf{h}|\mathbf{M})$ with mean $\hat{\bolds{\mu}}$ and variance--covariance matrix $\sigma^2\hat{\Sigma}$, then the linear predictors (\ref{point.est.untyped}) and (\ref{point.est.typed}) minimize
the integrated risk (assuming squared error loss) among all linear predictors [\citet{westharrison}]. In this sense they are the
best linear predictors, and so we refer to this method of imputation as Best Linear IMPutation or
BLIMP.\label{R1}
\end{Rem*}
\subsection{Extension to imputing genotype frequencies}

The development above considers imputing unobserved allele frequencies. In some settings one might also want to impute genotype
frequencies. A simple way to do this is to use an assumption of Hardy--Weinberg equilibrium: that is, to assume that if $y$ is
the allele frequency at the untyped SNP, then the three genotypes have frequencies $(1-y)^2, 2y(1-y)$ and $y^2$. Under our normal
model, the expected values of these three quantities can be computed:
\begin{eqnarray}
\mathrm{E}\bigl((1-y)^2|\mathbf{y}_t^{\mathrm{obs}}, \mathbf{M}\bigr) & = &\bigl(1 - \mathrm{E}( y|\mathbf{y}_t^{\mathrm{obs}},\mathbf{M})\bigr)^2 + \operatorname{Var}(y|\mathbf{y}_t^{\mathrm{obs}},\mathbf{M}),\nonumber\\
\mathrm{E}(y^2|\mathbf{y}_t^{\mathrm{obs}},\mathbf{M}) &=& (\mathrm{E}( y|\mathbf{y}_t^{\mathrm{obs}},\mathbf{M}) )^2 + \operatorname{Var}(y|\mathbf{y}_t^{\mathrm{obs}},\mathbf{M}),\\
\quad \mathrm{E}\bigl(2y(1-y)|\mathbf{y}_t^{\mathrm{obs}},\mathbf{M}\bigr) &=& 1 -\mathrm{E}\bigl((1-y)^2|\mathbf{y}_t^{\mathrm{obs}},\mathbf{M}\bigr)-\mathrm{E}(y^2|\mathbf{y}_t^{\mathrm{obs}},\mathbf{M}),\nonumber
\end{eqnarray}
where $\mathrm{E}( y|\mathbf{y}_t^{\mathrm{obs}},\mathbf{M})$ and $\operatorname{Var}(y|\mathbf{y}_t^{\mathrm{obs}},\mathbf{M})$ are given in (\ref{improvm}). These expectations can
be used as estimates of the unobserved genotype frequencies.

The method above uses only \textit{allele} frequency data at typed SNPs. If data are also available on \textit{genotype} frequencies, as might
be the case if the data are summary data from a regular genome scan in which all individuals were individually genotypes, then an alternative
approach that does not assume HWE is possible. In brief, we write the unobserved genotype frequencies as means of the genotype indicators,
$\indicator{g=0}$ and $\indicator{g=2}$ [analogous to expression (\ref{jointf})], and then derive expressions for the means and covariances
of these indicators both within SNPs and across SNPs. Imputation can then be performed by computing the appropriate conditional distributions
using the joint normal assumption, as in (\ref{basicm_sol}). See Appendix~\ref{gf.appx} for more details.

In practice, we have found these two methods give similar average accuracy (results not shown), although this could be because our data conform
well to Hardy--Weinberg equilibrium.

\subsection{Individual-level genotype imputation}

Although we developed the above model to tackle the imputation problem when individual genotypes are not available, it can also be applied
to the problem of individual-level genotype imputation when individual-level data \textit{are} available, by treating each individual as a pool
of two haplotypes (application of these methods to small pool sizes is justified by the \hyperref[R1]{Remark} above). For example, doubling
(\ref{point.est.untyped}) provides a natural estimate of the posterior mean genotype for an untyped SNP. For many applications this posterior
mean genotype may suffice; see \citet{guanstephens} for the use of such posterior means in downstream association analyses. If an estimate
that takes a value in \{$0,1,2$\} is desired, then a simple ad hoc procedure that we have found works well in practice is to round the
posterior mean to the nearest integer. Alternatively, a full posterior distribution on the three possible genotypes can be computed by using
the genotypic version of our approach (Appendix \ref{gf.appx}).

\subsection{Using unphased genotype panel data}

Our method can be readily adapted to settings where the panel data are unphased. To do this, we note that the estimates (\ref{basicm_mean}),
(\ref{basicm_var}) for $\bolds{\mu}$ and $\Sigma$ depend on the panel data only through the empirical mean and variance--covariance matrix of the panel
haplotypes. When the panel data are unphased, we simply replace these with 0.5 times the empirical mean and variance--covariance matrix of the
panel genotypes [since, assuming random mating, genotypes are expected to have twice the mean and twice the (co)variance of haplotypes]; see
\citet{weir79} for related discussion.

\subsection{Imputation without a panel}
In some settings, it may be desired to impute missing genotypes in a sample where no individuals are typed at all SNPs
(i.e., there is no panel $\mathbf{M}$), and each individual is typed at a different subset of SNPs. For example, this may arise if many individuals are sequenced at low coverage, as in the
currently-ongoing 1000 genomes project (\url{http://www.1000genomes.org}).
In the absence of a panel, we cannot directly obtain the mean and variance--covariance estimates $\hat{\bolds{\mu}}$ and $\hat{\Sigma}$ as in (\ref{basicm_mean}) and (\ref{basicm_var}). An alternative way to obtain these estimates is to treat each individual
genotype vector as a random sample from multivariate normal distribution $\mathrm{N}_p(\hat{\bolds{\mu}},\hat{\Sigma})$, and apply the ECM algorithm [\citet{mengrubin}] to perform maximum likelihood estimation. However, this approach does not incorporate shrinkage.
We therefore modify the algorithm in an \textit{ad-hoc} way to incorporate shrinkage in the conditional maximization step. See Appendix~\ref{ecm.appx} for details.

\section{Data application and results}

We evaluate methods described above by applying them to data from a subset of the WTCCC Birth Cohort, consisting of 1376 unrelated
British individuals genotyped on the Affymetrix 500K platform [\citet{wellcometrust}]. For demonstration purposes, we use only the
4329 SNPs from chromosome 22. We impute data at these SNPs using the 60 unrelated HapMap CEU parents [\citet{hapmap}] as the panel.
For the recombination parameters required in (\ref{covm}), we use the estimates distributed in the software package IMPUTE v1
[\citet{marchinietal}], which were estimated from the same panel using the software package PHASE [\citet{stephensscheet}].

In our evaluations we consider three types of applications: frequency imputation using summary-level data, individual-level genotype
imputation and noise reduction in DNA pooling experiments. We examine both the accuracy of point estimates and calibration of the
credible intervals.

\subsection{Frequency imputation using summary-level data}

In this section we evaluate the performance of (\ref{point.est.untyped}) for imputing frequencies at untyped SNPs. The observed data
consist of the marginal allele frequencies at each SNP, which we compute from the WTCCC individual-level genotype data. To assess imputation
accuracy, we perform the following cross-validation procedure: we mask the observed data at every 25th SNP, then treat the remaining SNPs as
typed and use them to impute the frequencies of masked SNPs and compare the imputation results with the actual observed frequencies. We repeat
this procedure 25 times by shifting the position of the first masked SNP. Because in this case the observed frequencies are obtained through
high quality individual-level genotype data,  we assume the experimental error parameter $\varepsilon^2 = 0$.

To provide a basis for comparison, we also perform the same experiment using the software package IMPUTE v1 [\citet{marchinietal}], which is
among the most accurate of existing methods for this problem. IMPUTE requires individual-level genotype data and outputs posterior genotype
probabilities for each unmeasured genotype. We therefore input the individual-level genotype data to IMPUTE and estimate the allele frequency
at each untyped SNP using the posterior expected frequency computed from the posterior genotype probabilities. Like our method, IMPUTE performs
imputation using the conditional distribution from Li and Stephens [\citet{listephens}]; however, it uses the full conditional distribution,
whereas our method uses an approximation based on the first two moments.  Furthermore, IMPUTE uses individual-level genotype data. For both
these reasons, we would expect IMPUTE to be more accurate than our method, and our aim is to assess how much we lose in accuracy by our
approximation and by using summary-level data.

To assess accuracy of estimated allele frequencies, we use the Root Mean Squared Error (RMSE),
\begin{equation}
  \mathrm{RMSE} = \sqrt{\frac{1}{J}\sum_{j=1}^{J} (y_{j} - \hat{y}_{j})^2},
\end{equation}
where $J$ is the number of SNPs tested (4329) and $y_{j}, \hat{y}_{j}$ are observed and imputed allele frequencies for SNP $j$ respectively.

The RMSE from our method was 0.0157 compared with 0.0154 from IMPUTE (Table \ref{genocomp}). Thus, for these data, using only summary-level
data sacrifices virtually nothing in accuracy of frequency imputation. Furthermore, we found that using an unphased panel (replacing the phased
HapMap CEU haplotypes with the corresponding unphased genotypes) resulted in only a very small decrease in imputation accuracy: $\mathrm{RMSE} = 0.0159$. In all cases the methods are substantially more accurate than a ``naive method'' that simply estimates the sample frequency using the panel frequency (Table \ref{genocomp}).

\begin{figure}

\includegraphics{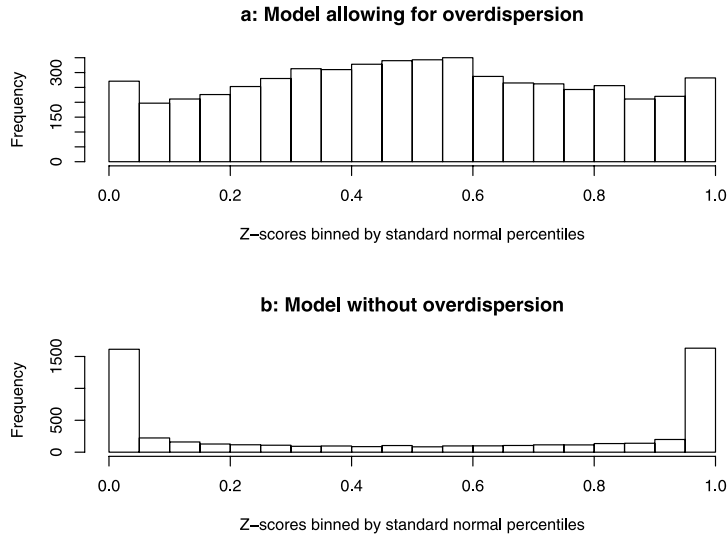}

\caption{Comparison of variance estimation in models with and
without overdispersions. The $Z$-scores are binned
according to the standard normal percentiles, for example, the first
bin (0 to 0.05) contains $Z$-score values from $-\infty$ to $-1.645$. If the
$Z$-scores are i.i.d. and strictly follow standard normal distribution,
we expect all the bins to have approximately equal height. }\label{caliplot}
\end{figure}

We also investigated the calibration of the estimated variances of the imputed frequencies from (\ref{improvm}). To do this, we constructed a $Z$-static for each test SNP $j$,
\begin{equation}
  Z_j = \frac{y_j-\mathrm{E}(y_j|\mathbf{y}^t,\mathbf{M})}{\sqrt{\operatorname{Var}(y_j |\mathbf{y}^t,\mathbf{M})}},
\end{equation}
where $y_j$ is the true observed frequency, and the conditional mean and variance are as in (\ref{improvm}). If the variances are
well calibrated, the $Z$-scores should follow a standard normal distribution (with slight dependence among $Z$-scores of neighboring
SNPs due to LD). Figure \ref{caliplot}a shows that, indeed, the empirical distribution of $Z$-scores is close to standard normal
(results are shown for phased panel; results for unphased panel are similar). Note that the overdispersion parameter plays a crucial
role in achieving this calibration. In particular, the $Z$-scores produced by the model without overdispersion (\ref{basicm}) do not
follow a standard normal distribution, with many more observations in the tails (Figure \ref{caliplot}b) indicating that the
variance is under-estimated.

\subsubsection{Comparison with unregularized linear frequency estimator}

\begin{figure}

\includegraphics{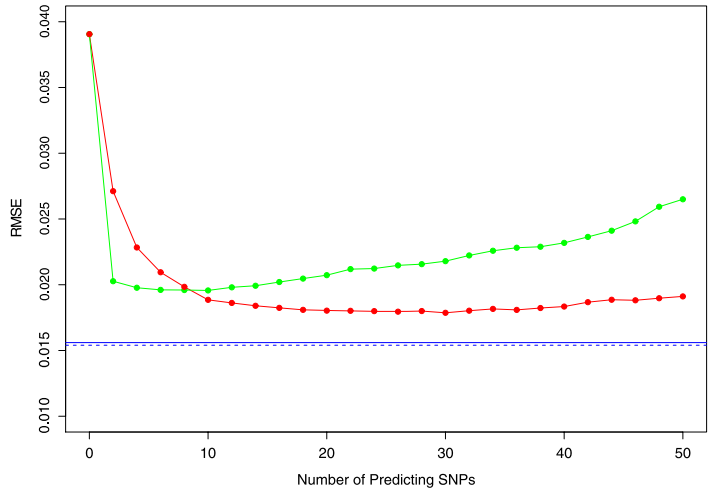}
\caption{Comparison between BLIMP
estimator and unregularized linear estimators. The lines show the RMSE
of each allele frequency estimator vs. number of predicting SNPs.
Results are shown for two schemes for selecting predicting SNPs:
flanking SNPs (red line) and correlated SNPs (green line). Neither
scheme is as accurate as BLIMP (blue solid line) or IMPUTE (blue dashed
line).}\label{regvswnd}
\end{figure}

To assess the role of regularization, we compared the accuracy of BLIMP with simple unregularized linear frequency
 estimators based on a small number of nearby ``predicting'' SNPs. (In fact, selecting a small number of SNPs can be viewed as
 a kind of regularization, but we refer to it as unregularized for convenience.) The unregularized linear estimator has the same form as
 in (\ref{basicm_sol}), but uses the unregularized estimates $\mathbf{f}^{\,\mathrm{panel}}$ and $\Sigma^{\mathrm{panel}}$ for $\hat{\bolds{\mu}}$
 and $\Sigma$.  We consider two schemes to select predictors: the first scheme selects $k$ flanking SNPs on either side of the target SNP (so $2k$
 predictors in total); the second scheme selects the $2k$ SNPs with the highest marginal correlation with the target SNP. Figure \ref{regvswnd} shows
 RMSE as predicting SNPs increases from 0 to 50. We find that the best performance of the unregularized methods is achieved by the first scheme, with
 a relatively large number of predicting SNPs (20--40); however, its RMSE is larger than that of IMPUTE and BLIMP.

\subsection{Individual-level genotype imputation}

Although very satisfactory\break methods already exist for individual-level genotype imputation, BLIMP has the potential advantage
of being extremely fast and low on memory-usage (see computational comparisons below). We therefore also assessed its performance
for individual-level genotype imputation. We used the same cross-validation procedure as in frequency imputation, but using individual-level
data as input. As above, we compared results from our approach with those obtained using IMPUTE v1.

We again use RMSE to measure accuracy of imputed (posterior mean) genotypes:
\begin{equation}
\mathrm{RMSE} = \sqrt{\frac{1}{m p}\sum_{j=1}^{p} \sum_{i=1}^{m} (g_{j}^i - \hat{g}_{j}^i)^2 },
\end{equation}
where $m$ is the number of the individuals (1376), $p$ is the total number of tested SNPs (4329) and
$g_{j}^i$, $\hat{g}_{j}^i$ are observed and estimated (posterior mean) genotypes for\vspace{1pt} individual $i$ at SNP $j$ respectively.

For comparison purposes, we also use a different measure of accuracy that is commonly used in this setting: the genotype error
rate, which is the number of wrongly imputed genotypes divide by the total number of imputed genotypes. To minimize the expected
value of this metric, one should use the posterior mode genotype as the estimated genotype. Thus, for IMPUTE v1 we used the
posterior mode genotype for this assessment with this metric. However, for simplicity, for our approach we used the posterior
mean genotype rounded to the nearest integer. (Obtaining posterior distributions on genotypes using our approach, as outlined
in Appendix \ref{gf.appx}, is considerably more complicated, and in fact produced slightly less accurate results, not shown.)

We found that under either metric BLIMP provides only very slightly less accurate genotype imputations than IMPUTE (Table \ref{genocomp}).
Further, as before, replacing the phased panel with an unphased panel produces only a small decrease in accuracy (Table \ref{genocomp}).
\begin{table}
\tablewidth=315pt
\caption{Comparison of accuracy of BLIMP and IMPUTE for frequency and
individual-level genotype imputations.
    The RMSE and error rate, defined in the text, provide different
    metrics for assessing accuracy; in all cases BLIMP was very
    slightly less accurate than IMPUTE. The ``naive method'' refers to
    the strategy of estimating the sample frequency of each untyped SNP by
    its observed frequency in the panel; this ignores information in the
    observed sample data, and provides a baseline level of accuracy
    against which the other methods can be compared}\label{genocomp}
     \begin{tabular*}{315pt}{@{\extracolsep{\fill}}lcc@{}}
\hline
      & \textbf{RMSE} & \textbf{Error rate} \\
       \hline
       \multicolumn{3}{@{}l@{}}{Frequency imputation}\\
       Naive method & 0.0397 & NA \\
       BLIMP (phased panel) & 0.0157 & NA  \\
       BLIMP (unphased panel) & 0.0159 & NA \\
       IMPUTE  & 0.0154 & NA \\[5pt]
      \multicolumn{3}{@{}l@{}}{Individual genotype imputation}\\
      BLIMP (phased panel) & 0.2339 & 6.46\%  \\
      BLIMP (unphased panel) & 0.2407 & 6.77\% \\
      IMPUTE  & 0.2303 & 6.30\% \\
      \hline
    \end{tabular*}
\end{table}

\begin{figure}

\includegraphics{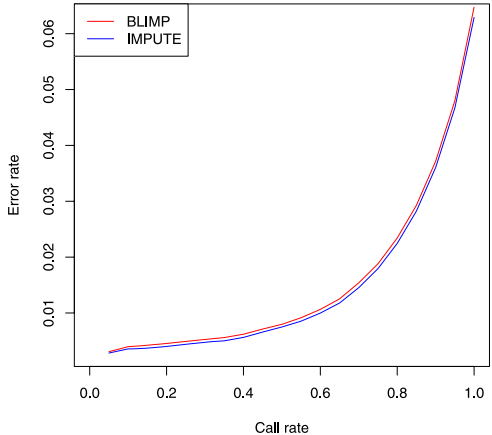}

\caption{Controlling individual-level genotype imputation error rate on
a per-SNP basis. For BLIMP, the error rate is
controlled by thresholding on the estimated variance for
imputed SNP frequencies; for IMPUTE the call threshold is determined by average maximum
posterior probability. These two different measures of per-SNP
imputation quality are strongly correlated (correlation coefficient${}=-0.983$). }\label{errcall}
\end{figure}

These results show average accuracy when all untyped SNPs are imputed. However, it has been observed previously
[e.g., \citet{marchinietal}, \citet{guanstephens}] that accuracy of calls at the most confident SNPs tends to be considerably higher
than the average. We checked that this is also true for BLIMP. To obtain estimates of the confidence of imputations at
each SNP, we first estimated $\sigma$ by maximum likelihood using the summary data across all individuals, and then computed the
variance for each SNP using (\ref{improvm}); note that this variance does not depend on the individual, only on the SNP. We then
considered performing imputation only at SNPs whose variance was less than some threshold, plotting the proportion of SNPs imputed
(``call rate'') against their average genotype error rate as this threshold varies. The resulting curve for BLIMP is almost identical
to the corresponding curve for IMPUTE (Figure \ref{errcall}).

\subsection{Individual-level genotype imputation without a panel}

We use the same WTCCC Birth Cohort data to assess our modified ECM algorithm for performing individual-level genotype imputation without using a panel. To create a data set with data missing at random, we mask each genotype, independently, with probability $m$. We create multiple data sets by varying $m$ from 5\% to 50\%. For each data set we run our ECM algorithm for 20 iterations. (Results using different starting points for the ECM algorithm were generally very consistent, and so results here are shown for a single starting point.)

We compare the imputation accuracy with the software package BIMBAM [\citet{guanstephens}] which implements
the algorithms from \citet{scheetstephens}.

BIMBAM requires the user to specify a number of ``clusters'' and other parameters related to the EM algorithm it uses: after experimenting with different settings we applied BIMBAM on each data set assuming $20$ clusters, with 10 different EM starting points, performing 20 iterations for each EM run. (These settings produced more accurate results than shorter runs.)

Overall, imputation accuracy of the two methods was similar (Table \ref{mar.cmp}), with BLIMP being slightly more accurate with larger amounts of missing data and \mbox{BIMBAM} being slightly more accurate for smaller amounts of missing data.

We note that, in this setting, some of the key computational advantages of our method are lost. In particular, when each individual is missing genotypes at different SNPs, one must effectively invert a different covariance matrix
for each individual. Furthermore, this inversion has to be performed multiple times, due to the iterative scheme.
For small amounts of missing data, the results from BLIMP we present here took less time than the results for BIMBAM, but for larger amounts the run times are similar.

\begin{table}
\tablewidth=220pt
\caption{Comparison of imputation error rates from BLIMP and BIMBAM for
individual genotype imputation without a panel}\label{mar.cmp}
\begin{tabular*}{220pt}{@{\extracolsep{\fill}}lcccc@{}}
\hline
&\multicolumn{4}{c@{}}{\textbf{Missing rate}} \\[-5pt]
&\multicolumn{4}{l@{}}{\hrulefill}\\
 &\textbf{5\%}&\textbf{10\%}&\textbf{20\%}&\textbf{50\%}\\
\hline
BIMBAM  & 5.79\% & 6.35\% & 7.15\% & 9.95\% \\
BLIMP ECM & 6.07\% & 6.49\% & 7.31\% & 9.91\% \\
\hline
\end{tabular*}
\end{table}

\subsection{Noise reduction in pooled experiment}

We used simulations to assess the potential for our approach to improve allele frequency estimates from noisy data in DNA pooling
experiments [equation (\ref{nrd})].
To generate noisy observed data, we took allele frequencies of 4329 genotyped SNPs from the WTCCC
Birth Cohort chromosome 22 data as true values, and added independent and identically distributed $N(0,\varepsilon^2)$ noise terms to each
true allele frequency. Real pooling data will have additional features not captured by these simple simulations (e.g., biases toward one of the alleles), but our aim here is simply
to illustrate the potential for methods like ours to reduce noise in this type of setting.
We varied $\varepsilon$ from 0.01 to 0.18 to  examine different noise levels. Actual noise levels in pooling
experiments will depend on technology and experimental protocol; to give a concrete example, \citet{meaburnetal} found differences
between allele frequency estimates from pooled genotyping and individual genotyping of the same individuals, at 26 SNPs, in the range
0.008--0.077 (mean 0.036).

We applied our method to the simulated data by first estimating $\sigma$ and $\varepsilon$ using (\ref{improvm}), and then, conditional on
these estimated parameters, estimating the allele frequency at each observed SNP using the posterior mean given in equation (\ref{nrd}).
We assessed the accuracy (RMSE) of these allele frequency estimates by comparing them with the known true values.

\begin{figure}

\includegraphics{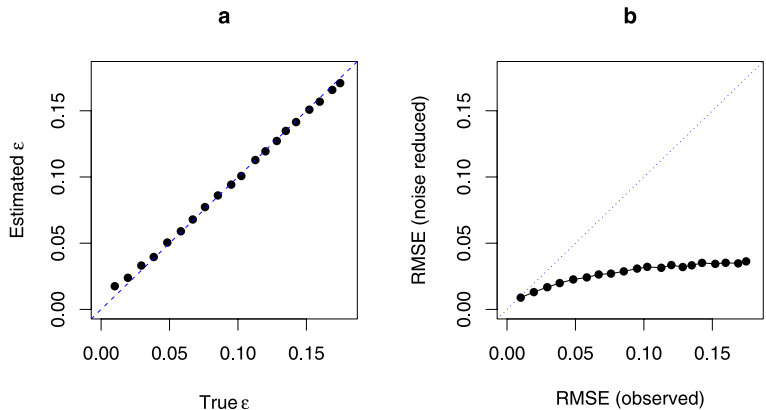}

\caption{(\textup{a}) Detection of experimental noise in simulated
data. The simulated data sets are generated by adding Gaussian noise
$N(0, \varepsilon^2)$ to the actual observed WTCCC frequencies. The estimated $\varepsilon$
values are plotted against the true $\varepsilon$ values used for
simulation. We estimate $\varepsilon$ using maximum likelihood by
(\protect\ref{obs.likelihood}). (\textup{b}) An illustration on the
effect of noise reduction in various noise levels.  RMSE from noise reduced estimates are
plotted against RMSE from direct noisy observations.  The noise reduced
frequency estimates are posterior means obtained from model (\protect\ref{nrd}).}\label{noise_rd}
\end{figure}

We found that our method was able to reliably estimate the amount of noise present in the data: the estimated values for the error parameter
$\varepsilon$ show good correspondence with the standard deviation used to simulate the data (Figure \ref{noise_rd}a), although for high noise
levels we underestimate the noise because some of the errors are absorbed by the parameter $\sigma$.

More importantly, we found our estimated allele frequency estimates were consistently more accurate than the direct (noisy) observations,
with the improvement being greatest for higher noise levels (Figure \ref{noise_rd}b). For example, with $\varepsilon = 0.05$ our method reduced
the RMSE by more than half, to 0.024.

\subsection{Computational efficiency}

Imputation using our implementation of\break BLIMP is highly computationally efficient. The computational efficiency is especially notable
when dealing with large panels:  the panel is used only to estimate $\muLS$ and $\SigmaLS$, which is both quick and done just once, after
which imputation computations do not depend on panel size. Our implementations also take advantage of the sparsity of $\SigmaLS$ to increase
running speed and reduce memory usage. To give a concrete indication of running speed, we applied BLIMP to the WTCCC Birth Cohort data on
chromosome 22, containing 4329 genotyped and 29,697 untyped Hapmap SNPs on 1376 individuals, using a Linux system with eight-core Intel
Xeon 2.66 GHz processors (although only one processor is used by our implementation). The running time is measured by `real' time reported
by the Unix ``time'' command. For frequency imputation, BLIMP took 9 minutes and 34 seconds, with peak memory usage of 162 megabytes; for
individual-level genotype imputation, BLIMP took 25 minutes, using under 300 megabytes of memory. As a comparison, IMPUTE v1 took 195 minutes
for individual-level genotype imputation, with memory usage exceeding 5.1 gigabytes.

Since these comparisons were done, we note that a new version of IMPUTE (v2) has been released [\citet{howieetal}]. This new version gives
similar imputation accuracy in the settings we described above; it runs more slowly than v1 but requires less memory.

\section{Conclusion and discussion}

Imputation has recently emerged as an important and powerful tool in genetic association studies. In this paper we propose a set of statistical
tools that help solve the following problems:
\begin{enumerate}
  \item Imputation of allele frequencies at untyped SNPs when only summary-level data are available at typed SNPs.
  \item Noise reduction for estimating allele frequencies from DNA pooling-based experiments.
  \item Fast and accurate individual-level genotype imputation.
\end{enumerate}
The proposed methods are simple, yet statistically elegant,  and computationally extremely efficient. For individual-level genotype imputation
the imputed genotypes from this approach are only very slightly less accurate than state-of-the-art methods. When only summary-level data are
available, we found that imputed allele frequencies were almost as accurate as when using full individual genotype data.

The linear predictor approach to imputation requires only an estimate of the mean and the covariance matrix among SNPs. Our approach to obtaining
these estimates is based on the conditional distribution from \citet{listephens}; however, it would certainly be possible
to consider other estimates, and, specifically, to use other approaches to shrink the off-diagonal terms in the covariance matrix.
An alternative, closely-related, approach is to obtain a linear predictor directly by training a linear regression to predict each SNP,
using the panel as a training set, and employing some kind of regularization scheme to solve potential problems with overfitting caused by
large $p$ small $n$. This approach has been used in the context of individual-level genotype imputation by A. Clark (personal communication)
and \citet{yuschaid}. However, the choice of appropriate regularization is not necessarily straightforward, and different regularization
schemes can provide different results [\citet{yuschaid}]. Our approach of regularizing the covariance matrix using the conditional distribution
from \citet{listephens} has the appeal that this conditional distribution has already been shown to be very effective for individual genotype
imputation, and for modeling patterns of correlation among SNPs more generally [\citet{listephens}, \citet{stephensscheet}, \citet{servinstephens}, \citet{marchinietal}].
Furthermore, the fact that, empirically, BLIMP's accuracy is almost as good as the best available purpose-built methods for this problem suggests
that alternative approaches to regularization are unlikely to yield considerable improvements in accuracy.

The accuracy with which linear combinations of typed SNPs can predict untyped SNPs is perhaps somewhat surprising. That said, theoretical
arguments for the use of linear combinations have been given in previous work. For example, \citet{claytonetal} showed by example that, when
SNP data are consistent with no recombination (as might be the case for markers very close together on the genome), each SNP can be written
as a linear regression on the other SNPs.  Conversely, it is easy to construct hypothetical examples where linear predictors would fail badly.
For example, consider the following example from \citet{nicolaemd}: 3 SNPs form 4 haplotypes, 111, 001, 100 and 010, each at frequency 0.25
in a population. Here the correlation between every pair of SNPs \mbox{is 0}, but knowing any 2 SNPs is sufficient to predict the third SNP precisely.
Linear predictors cannot capture this ``higher order'' interaction information,  so produce suboptimal imputations in this situation. In contrast,
other methods (including \mbox{IMPUTE}) could use the higher-order information to produce perfect imputations. The fact that, empirically, the linear
predictor works well suggests that this kind of situation is rare in real human population genotype data. Indeed, this is not so surprising when
one considers that, from population genetics theory, SNPs tend to be uncorrelated only when there is sufficiently high recombination rate between
them, and recombination will tend to break down any higher-order correlations as well as pairwise correlations.

Besides HMM-based methods, another type of approach to genotype imputation that has been proposed is to use ``multi-marker'' tagging
[\citet{debakkeretal}, \citet{plink}, \citet{nicolae}]. A common feature of these methods is to preselect a (relatively small) \textit{subset} of
``tagging''
SNPs or haplotypes from all typed SNPs based on some LD measure threshold, and then use a possibly nonlinear approach to predicting
untyped SNPs from this subset. Thus, compared with our approach, these methods generally use a more complex prediction method based on a
smaller number of SNPs. Although we have not compared directly with these methods here,  published comparisons [\citet{howieetal}] suggest
that they are generally noticeably less accurate than HMM-based methods like IMPUTE, and, thus, by implication, less accurate than BLIMP.
That is, it seems from these results that, in terms of average accuracy, it is more important to make effective use of low-order correlations
from all available SNPs that are correlated with the target untyped SNP than to take account of unusual higher-order correlations that may
occasionally exist.

Our focus here has been on the accuracy with which untyped SNP allele frequencies can be imputed. In practice, an important application
of these imputation methods is to test untyped alleles for association with an outcome variable (e.g., case-control status). Because our
allele frequency predictors are linear combinations of typed SNP frequencies, each test of an untyped SNP is essentially a test for
differences between a given linear combination of typed SNPs in case and control groups. Several approaches to this are possible; for example,
the approach in \citet{nicolae} could be readily applied in this setting. The resulting test would be similar to the test suggested in
\citet{homeretalpooling} which also uses a linear combination of allele frequencies at typed SNPs to test untyped SNPs for association with
case-control status in a pooling context. The main difference is that their proposed linear combinations are {ad hoc}, rather than
being chosen to be the best linear imputations; as such, we expect that appropriate use of our linear imputations should result in more
powerful tests, although a demonstration of this lies outside the scope of this paper.

\begin{appendix}

\section{Learning from panel using the\break Li and Stephens model}\label{ln.appx}
In this section we show the calculation of $\hat{\mv}$ and $\hat{\Sigma}$ using phased population panel data. Following the Li and Stephens model,
we assume that there are $K$ template haplotypes in the panel. For a new haplotype $\hv$ sampled from the same population, it can be modeled as
an imperfect mosaic of existing template haplotypes in the panel. Letting $\ev_j$ denote the $j$th unit vector in $K$ dimensions (1 in the $j$th coordinate,
$0$'s elsewhere), we define random vector $\Zv_{t}$ to be $\ev_j$ if haplotype $\hv$ at locus $t$ copies from $j$th template haplotype. The model
also assumes $\Zv_{1},\ldots,\Zv_{n}$ form a Markov chain in state space $\{ \ev_1,\ldots,\ev_K \} $ with transition probabilities
\begin{equation}
  \operatorname{Pr}( \Zv_{t} = \ev_m | \Zv_{t-1} = \ev_n, \Mv ) = (1-r_t) \indicator{\ev_m = \ev_n} + r_t {\mathbf{e}^{\prime}}_m \av,
\end{equation}
where $\av = \frac{1}{K} \cdot {\mathbf 1}$ and $r_t=1-\exp(-\rho_t/K)$ is a parameter that controls the probability that $\hv$ switches copying template at locus $t$. The initial-state
probabilities of the Markov chain are given by
\begin{equation}
  \pi( \Zv_{1} = \ev_k | \Mv ) = \frac{1}{K}\qquad\mbox{for }k=1,\ldots,K .
\end{equation}
It is easy to check that the initial distribution $\pi$ is also the stationary distribution of the described Markov chain. Because
the chain is initiated at the stationary state, it follows that, conditional on $\Mv$,
\begin{equation}
  \Zv_1 =^d \Zv_2 =^d \cdots =^d \Zv_p =^d \pi.
\end{equation}
Therefore, marginally, the means and variances of $\Zv_t$'s have the following simple forms:
\begin{eqnarray}
  \mathrm{E}(\Zv_1|\Mv)&=&\cdots=\mathrm{E}(\Zv_p|\Mv)=\av, \\
  \operatorname{Var}(\Zv_1|\Mv)&=&\cdots=\operatorname{Var}(\Zv_n|\Mv)=\operatorname{diag}(\av)-\av\av^\prime.
\end{eqnarray}
Let $K$-dimensional vector $\qv_t^{\mathrm{panel}}$ denote the binary allelic state of panel haplotypes at locus $t$ and scalar parameter $\theta$ represents mutation. The emission distribution in the Li and Stephens model is given by
\begin{equation}
\operatorname{Pr}(h_{t}=1|\Zv_{t}=\ev_k,\Mv) = (1-\theta)\ev_k^{\prime}\qv_t^{\mathrm{panel}} + \tfrac{1}{2}\theta,
\end{equation}
that is, with probability $1-\theta$, $\hv$ perfectly copies from the $k$th template in the panel at locus $t$, while, with probability $\theta$, a mutation occurs and $h_t$  ``mutates'' to allele 0 or 1 equally likely. If we define $\pv_t =  (1-\theta)\qv_t^{\mathrm{panel}} + \frac{\theta}{2}\mathbf{1}$,
then the emission distribution can be written as
\begin{equation}
\operatorname{Pr}(h_{t}=1|\Zv_{t},\Mv ) = \mathrm{E}(h_{t} |\Zv_{t},\Mv ) = \pv_t^{\prime} \Zv_t.
\end{equation}
The goal here is to find the closed-form representations of the first two moments of joint distribution $( h_1, h_2,\ldots, h_p )$ given the observed template panel $\Mv$. For marginal mean and variance of $h_t$, it follows that
\begin{eqnarray}
  \label{ehm}
      \mathrm{E}(h_t|\Mv) &=& \mathrm{E}(\mathrm{E}(h_t|\Zv_t,\Mv)|\Mv)\nonumber\\
    &=& \pv_t^\prime \mathrm{E}(\Zv_t|\Mv)\\
    &=&  (1-\theta)\cdot f_t^{\mathrm{panel}}+\frac{\theta}{2},\nonumber\\
    \operatorname{Var}(h_t|\Mv)&=& (1-\theta)^2 f_t^{\mathrm{panel}} (1 - f_t^{\mathrm{panel}}) + \frac{\theta}{2}\biggl(1-\frac{\theta}{2}\biggr),
    \end{eqnarray}
where $f_t^{\mathrm{panel}} = \qv_t^\prime \cdot \av $ is the observed allele frequency at locus $t$ from panel $\Mv$.
Finally, to compute $\operatorname{Cov}(h_s,h_t)$ for some loci $s < t$, we notice that, conditional on $\Zv_s$ and $\Mv$, $h_s$ and $h_t$
are independent and
\begin{eqnarray}
      \mathrm{E}( h_s \cdot h_t|\Mv) &= &\mathrm{E}\bigl(\mathrm{E}(h_s \cdot h_t|\Zv_s,\Mv)|\Mv\bigr)\nonumber
      \\[-8pt]\\[-8pt]
    &=& \mathrm{E}\bigl(\mathrm{E}(h_s|\Zv_s,\Mv)\cdot\mathrm{E} (h_t|\Zv_s,\Mv)\vert\Mv\bigr)\nonumber.
 \end{eqnarray}
Let $r_{st}$ denote the switching probability between $s$ and $t$, and $\mathrm{E} (h_t|\Zv_s,\Mv)$ can be calculated from
\begin{eqnarray}
      \mathrm{E}(h_t|\Zv_s,\Mv)&=&\mathrm{E}(\mathrm{E}(h_t|\Zv_t,\Zv_s,\Mv)|\Zv_s,\Mv)\nonumber \\
    &=&\mathrm{E}( \Zv_t^\prime \pv_t | \Zv_s,\Mv) \\
    &=&\bigl((1-r_{st})\Zv_s^\prime + r_{st} \av^\prime\bigr) \pv_t.\nonumber
 \end{eqnarray}
Therefore,
\begin{eqnarray}
  \label{ehhm}
    \qquad\mathrm{E}( h_s \cdot h_t | \Mv ) &=&\pv_s^\prime{\mathrm{E}}\bigl(\Zv_s \bigl((1-r_{st})\Zv_s^\prime + r_{st} \av^\prime\bigr)|\Mv\bigr) \pv_t\nonumber\\
    &=&(1-r_{st}) \pv_s^\prime {\operatorname{Var}}(\Zv_s) \pv_t + \pv_s^\prime \av \av^\prime \pv_t \\
    &=&(1-r_{st}) \cdot \pv_s^\prime \bigl(\operatorname{diag}(\av) - \av \av^\prime\bigr) \pv_t +\mathrm{E}(h_s|\Mv)\mathrm{E}(h_t|\Mv).\nonumber
\end{eqnarray}
It follows that
\begin{equation}
  \operatorname{Cov}(h_s, h_t|\Mv) = (1-\theta)^2(1-r_{st})(f_{st}^{\mathrm{panel}} - f_s^{\mathrm{panel}} f_t^{\mathrm{panel}}),
\end{equation}
where $f_{st}^{\mathrm{panel}}$ is the panel frequency of the haplotype ``$1$--$1$'' consisting of loci $s$ and $t$.

In conclusion, under the Li and Stephens model, the distribution $\hv\vert\Mv$ has expectation
\begin{equation}
  \hat{\mv}= \mathrm{E}(\hv|\Mv) = (1-\theta)\fv^{\,\mathrm{panel}} +\frac{\theta}{2}\mathbf{1},
\end{equation}
where $\fv^{\,\mathrm{panel}}$ is the $p$-vector of observed frequencies of all $p$ SNPs in the panel, and variance
\begin{equation}
  \hat \Sigma = \operatorname{Var}(\hv|\Mv) =  (1-\theta)^2 S + \frac{\theta}{2}\biggl(1-\frac{\theta}{2}\biggr) I,
\end{equation}
where matrix $S$ has the structure
\begin{equation}
  S_{ij} = \cases{f_{i}^{\mathrm{panel}}(1-f_{i}^{\mathrm{panel}}),&\quad$i=j$,
  \cr(1 - r_{ij})(f_{ij}^{\mathrm{panel}} - f_i^{\mathrm {panel}} f_j^{\mathrm{panel}}),&\quad$i \ne j$.}
\end{equation}

\section{Derivation of joint genotype frequency distribution}\label{gf.appx}

In this section we derive the joint genotype frequency distribution based on the Li and Stephens model.

Let $g_{it}$ denote the genotype of individual $i$ at locus $t$. The sample frequency of genotype $0$ at locus $t$ is given by
\begin{equation}
 \operatorname{Pr}(g_t=0) = p_0^{g_t} = \frac{1}{n}\sum_{i=1}^{n}\indicator{g_{it}=0}.
\end{equation}
Similarly, genotype frequencies $p_1^{g_t} = \operatorname{Pr}(g_t=1)$ and $p_2^{g_t}=\operatorname{Pr}(g_t=2)$ can be obtained by averaging indicators $\indicator{g_{it}=1}$ and $\indicator{g_{it}=2}$ over the samples respectively. Because of the restriction
\begin{equation}
 \indicator{g_{it}=0} +  \indicator{g_{it}=1} +  \indicator{g_{it}=2} = 1,
\end{equation}
given any two of the three indicators, the third one is uniquely determined. Let $\gv_i$ denote $2p$-vector $(\indicator{g_{i1}=0},\indicator{g_{i1}=2},\ldots,  \indicator{g_{ip}=0},\indicator{g_{ip}=2})$ and
\begin{equation}
     \yvg = (p_0^{g_1}p_2^{g_1}\cdots p_0^{g_p}p_2^{g_p})^\prime = \frac{1}{n}\sum_{i=1}^n \gv_i.
\end{equation}
Assuming that $\gv_1,\ldots,\gv_n$ are i.i.d. draws from conditional distribution $\operatorname{Pr}(\gv|\Mv)$, by the central limit theorem as sample size $n$ is large, it follows
\begin{equation}
     \yvg|\Mv\sim\mathrm{N}_{2p}(\mvg, \Sigma_g),
\end{equation}
where $\mvg = \mathrm{E}(\gv|\Mv)$ and $\Sigma_g = \operatorname{Var}(\gv|\Mv)$.

For the remaining part of this section, we derive the closed-form expressions for $\mvg$ and $\Sigma_g$ based on the Li and Stephens model.

Let $\hv^a$ and $\hv^b$ denote the two composing haplotypes for some genotype sampled from the population. Note that
\begin{eqnarray}
    \indicator{g_t = 0} &=& (1-h_t^a)(1-h_t^b), \nonumber\\[-8pt]\\[-8pt]
    \indicator{g_t = 2} &=& h_t^a h_t^b.\nonumber
\end{eqnarray}
Given panel $\Mv$, the two composing haplotypes are also assumed to be independent and identically distributed. Following the results from Appendix \ref{ln.appx}, we obtain that
\begin{eqnarray}
      \mathrm{E}\bigl(\indicator{g_t = 0}|\Mv\bigr) &=& \bigl(1-\mathrm{E}(h_t|\Mv)\bigr)^2,\nonumber \\
    \mathrm{E}\bigl(\indicator{g_t = 2}|\Mv\bigr) &=& \mathrm{E}(h_t|\Mv)^2, \nonumber\\
    \operatorname{Var}\bigl(\indicator{g_t = 0}|\Mv\bigr) &=&\bigl(1-\mathrm{E}(h_t|\Mv)\bigr)^2 \cdot \bigl( 1 - \bigl(1-\mathrm{E}(h_t|\Mv)\bigr)^2\bigr),\\
    \operatorname{Var}\bigl(\indicator{g_t = 2}|\Mv\bigr) &=& \mathrm{E}(h_t|\Mv)^2 \cdot \bigl( 1 - \mathrm{E}(h_t|\Mv)^2\bigr),\nonumber\\
    \quad\operatorname{Cov}\bigl(\indicator{g_t = 0},\indicator{g_t = 2}|\Mv\bigr) &=& -\bigl(1-\mathrm{E}(h_t|\Mv)\bigr)^2\cdot\mathrm{E}(h_t|\Mv)^2,\nonumber
  \end{eqnarray}
where $\mathrm{E}(h_t|\Mv)$ is given by (\ref{ehm}).

To compute covariance across different loci $s$ and $t$, we note that
 \begin{eqnarray}
  \indicator{g_s = 0} \indicator{g_t = 0} &=& (1-h_s^a)(1-h_s^b)(1-h_t^a)(1-h_t^b) ,\nonumber\\
  \indicator{g_s = 2} \indicator{g_t = 2} &=& h_s^a h_s^b h_t^a h_t^b,\nonumber\\[-8pt]\\[-8pt]
  \indicator{g_s = 0} \indicator{g_t = 2} &=& (1-h_s^a)(1-h_s^b) h_t^a h_t^b ,\nonumber\\
  \indicator{g_s = 2} \indicator{g_t = 0} &=& h_s^a h_s^b (1-h_t^a)(1-h_t^b).\nonumber
  \end{eqnarray}
Then all the covariance terms across different loci can be represented using $\mathrm{E}(h_s h_t | \Mv)$, which is given by (\ref{ehhm}). For example,
\begin{eqnarray}
&&\operatorname{Cov}\bigl(\indicator{g_s = 2},\indicator{g_t=2}|\Mv\bigr)\nonumber\\[-8pt]\\[-8pt]
  &&\qquad=\operatorname{Cov}(h_s,h_t|\Mv)^2\nonumber + 2 \mathrm{E}(h_s|\Mv) \cdot \mathrm{E}(h_t|\Mv) \cdot\operatorname{Cov}(h_s,h_t|\Mv).\nonumber
\end{eqnarray}

\section{Modified ECM algorithm for imputing genotypes without a panel}\label{ecm.appx}

In this section we show our modified ECM algorithm for genotype imputation without a panel.

By our assumption,  each individual genotype $p$-vector $\gv^i$ is a random sample\vspace{1pt} from the multivariate normal distribution $\mathrm{N}_p(\muLS,\SigmaLS)$, and different individual vectors may have different missing entries. Suppose we have $n$ individual samples, then let $\Gv_{\mathrm{obs}}$ denote the set of all typed genotypes across all individuals and $\gv^i_{\mathrm{obs}}$ denote the typed genotypes for individual $i$.

In the E step of the ECM algorithm, we compute the expected values of the sufficient statistics $\sum_{i=1}^n g_j^i$ for $j = 1,\ldots,p$ and
$\sum_{i=1}^n g_j^i g_k^i$ for $j,k = 1,\ldots,p$ conditional on $\Gv_{\mathrm{obs}}$ and the current estimate for $(\muLS,\SigmaLS)$. Specifically,
in the $t$th iteration,
\begin{eqnarray}
\mathrm{E}\Biggl(\sum_{i=1}^n g_j^i\Bigl|\Gv_{\mathrm{obs}}, \muLS^{(t)},\SigmaLS^{(t)}\Biggr) &=& \sum_{i=1}^n g_j^{i,(t)}, \\
\mathrm{E}\Biggl(\sum_{i=1}^n g_j^i g_k^i\Bigl|\Gv_{\mathrm{obs}}, \muLS^{(t)},\SigmaLS^{(t)}\Biggr) &=& \sum_{i=1}^n \bigl( g_j^{i,(t)} g_k^{i,(t)} + c_{jk}^{i,(t)}\bigr),
\end{eqnarray}
where
\begin{equation}
  g_j^{i,(t)} = \cases{g_j^i,&\quad\mbox{if $g_j^i$ is typed},
  \cr\mathrm{E}\bigl(g_j^i | \gv_{\mathrm{obs}}^i, \muLS^{(t)},\SigmaLS^{(t)}\bigr),&\quad\mbox{if $g_j^i$ is untyped,}}
\end{equation}
and
\begin{equation}
 \qquad c_{jk}^{i,(t)}= \cases{0,&\quad\mbox{if $g_j^i$ or $g_k^i$ is typed},
  \cr\operatorname{Cov}\bigl(g_j^i,g_k^i | \gv_{\mathrm{obs}}^i, \muLS^{(t)},\SigmaLS^{(t)}\bigr),&\quad\mbox{if $g_j^i$ and $g_k^i$ are both
  untyped.}}\hspace{-6pt}
\end{equation}
The calculation of $\mathrm{E}(g_j^i | \gv_{\mathrm{obs}}^i, \muLS^{(t)},\SigmaLS^{(t)})$ and $\operatorname{Cov}(g_j^i,g_k^i | \gv_{\mathrm{obs}}^i, \muLS^{(t)},\SigmaLS^{(t)})$ follows directly from (\ref{basicm_sol}).

In the conditional maximization step, we first update the estimates for $\fvPanel$ and $\SigmaPanel$ sequentially, that is,
\begin{equation}
  f_j^{\mathrm{panel},(t+1)} = \frac{1}{n} \mathrm{E}\Biggl(\sum_{i=1}^n g_j^i\Bigl|\Gv_{\mathrm{obs}}, \muLS^{(t)}\Biggr)\qquad\mbox{for $j=1,\ldots,p$,}
\end{equation}
and
\begin{eqnarray}
    \Sigma_{jk}^{\mathrm{panel},(t+1)} &=& \frac{1}{n}\mathrm{E}\Biggl(\sum_{i=1}^n g_j^i g_k^i\Bigl|\Gv_{\mathrm{obs}},
    \muLS^{(t)},\SigmaLS^{(t)}\Biggr)\nonumber\\[-8pt]\\[-8pt]
                                    &&{}-f_j^{\mathrm{panel},(t+1)}f_k^{\mathrm{panel},(t+1)}\qquad\mbox{for $j,k=1,\ldots,p$.}\nonumber
 \end{eqnarray}

Finally, we update the shrinkage estimates $\muLS$ and $\SigmaLS$ using
\begin{eqnarray}
  \muLS^{(t+1)} &=& (1-\theta)\fv^{\,\mathrm{panel},(t+1)} +\frac{\theta}{2}{\mathbf 1}, \\
  \SigmaLS^{(t+1)} &=& (1-\theta)^2 S^{(t+1)} + \frac{\theta}{2}\biggl(1-\frac{\theta}{2}\biggr) I,
\end{eqnarray}
where
\begin{equation}
 S_{jk}^{(t+1)} = \cases{\Sigma_{jk}^{\mathrm{panel},(t+1)},&\quad$j=k$,
  \cr\displaystyle\exp\biggl(-\frac{\rho_{jk}}{2n}\biggr) \Sigma_{jk}^{\mathrm{panel},(t+1)},&\quad$ij \ne k$.}
  \end{equation}

We initiated the ECM algorithm by setting  $\fv^{\,\mathrm{panel},(0)}$ to the marginal means from all observed data and $ \Sigma^{\mathrm{panel},(0)}$ to a diagonal matrix with diagonal entries being empirical variance computed from typed SNPs.

\end{appendix}
\section*{Acknowledgments}

We thank Yongtao Guan and Bryan Howie for helpful discussions.

\printaddresses


\begin{thebibliography}{}
\bibitem[\protect\citeauthoryear{Browning and Browning}{2007}]{browning}
\textsc{Browning, S.} and \textsc{Browning, B.} (2007).
Rapid and accurate haplotype phasing and missing data inference for whole genome association studies using localized haplotype clustering.
\textit{American Journal of Human Genetics} \textbf{81} 1084--1097.

\bibitem[\protect\citeauthoryear{Clayton, Chapman and Cooper}{2004}]{claytonetal}
\textsc{Clayton, D., Chapman, J.} and \textsc{Cooper, J.} (2004).
Use of unphased multilocus genotype data in indirect association studies.
\textit{Genetic Epidemiology} \textbf{27} 415--428.

\bibitem[\protect\citeauthoryear{de~Bakker et~al.}{2005}]{debakkeretal}
\textsc{de~Bakker, P., Yelensky, R., Pe'er, I., Gabriel, S., Daly, M.} and \textsc{Altshuler, D.} (2005).
Efficiency and power in genetic association studies.
\textit{Nature Genetics} \textbf{37} 1217--1223.

\bibitem[\protect\citeauthoryear{Guan and Stephens}{2008}]{guanstephens}
\textsc{Guan, Y.} and \textsc{Stephens, M.} (2008).
Practical issues in imputation-based association mapping.
\textit{P{L}o{S} Genetics} \textbf{4} e1000279.

\bibitem[\protect\citeauthoryear{Homer et~al.}{2008a}]{homeretalpooling}
\textsc{Homer, N., Tembe, W., Szelinger, S., Redman, M., Stephan, D., Pearson, J., Nelson, D.} and \textsc{Craig, D.} (2008a).
Multimarker analysis and imputation of multiple platform pooling-based genome-wide association studies.
\textit{Bioinformatics} \textbf{24} 1896--1902.

\bibitem[\protect\citeauthoryear{Homer et~al.}{2008b}]{homeretalmixture}
\textsc{Homer, N., Szelinger, S., Redman, M., Duggan, D., Tembe, W., Muehling, J., Pearson, J., Stephan, D., Nelson, S.} and \textsc{Craig, D.} (2008b).
Resolving individuals contributing trace amounts of dna to highly complex mixtures using high-density snp genotyping microarrays.
\textit{P{L}o{S} Genetics} \textbf{4} e1000167.

\bibitem[\protect\citeauthoryear{Howie, Donnelly and Marchini}{2009}]{howieetal}
\textsc{Howie, B., Donnelly, P.} and \textsc{Marchini, J.} (2009).
A flexible and accurate genotype imputation method for the next generation of genome-wide association studies.
\textit{P{L}o{S} Genetics} \textbf{5} e1000529.

\bibitem[\protect\citeauthoryear{Huang et~al.}{2009}]{huangetal}
\textsc{Huang, L., Li, Y., Singleton, A., Hardy, J., Abecasis, G., Rosenberg, N.} and \textsc{Scheet, P.} (2009).
Genotype-imputation accuracy across worldwide human populations.
\textit{American Journal of Human Genetics} \textbf{84} 235--250.

\bibitem[\protect\citeauthoryear{Hudson}{2001}]{hudson01}
\textsc{Hudson, R.} (2001).
Two-locus sampling distributions and their application.
\textit{Genetics} \textbf{159} 1805--1817.

\bibitem[\protect\citeauthoryear{Li and Stephens}{2003}]{listephens}
\textsc{Li, N.} and \textsc{Stephens, M.} (2003).
Modelling linkage disequilibrium and identifying recombination hotspots using snp data.
\textit{Genetics} \textbf{165} 2213--2233.

\bibitem[\protect\citeauthoryear{Li, Ding and Abecasis}{2006}]{abecasis}
\textsc{Li, Y., Ding, J.} and \textsc{Abecasis, G.} (2006).
Mach 1.0: Rapid haplotype reconstruction and missing genotype inference.
\textit{American Journal of Human Genetics} \textbf{79} S2290.

\bibitem[\protect\citeauthoryear{Marchini et~al.}{2007}]{marchinietal}
\textsc{Marchini, J., Howie, B., Myers, S., McVean, G.} and \textsc{Donnelly, P.} (2007).
A new multipoint method for genome-wide association studies by imputation of genotypes.
\textit{Nature Genetics} \textbf{39} 906--913.

\bibitem[\protect\citeauthoryear{McCullagh and Nelder}{1989}]{mccullaghnelder}
\textsc{McCullagh, P.} and \textsc{Nelder, J.} (1989).
\textit{Generalized Linear Models}, 2nd ed. Chapman and Hall, London.
\MR{0727836}

\bibitem[\protect\citeauthoryear{McVean, Awadalla and Fearnhead}{2002}]{mcveanetal}
\textsc{McVean, G., Awadalla, P.} and \textsc{Fearnhead, P.} (2002).
A coalescent-based method for detecting and estimating recombination from gene sequences.
\textit{Genetics} \textbf{160} 1231--1241.

\bibitem[\protect\citeauthoryear{Meaburn et~al.}{2006}]{meaburnetal}
\textsc{Meaburn, E., Butcher, L., Schalkwyk, L.} and \textsc{Plomin, R.} (2006).
Genotyping pooled DNA using 100k snp microarrays: A step towards genomewide association scans.
\textit{Nucleic Acids Research} \textbf{34} e28.

\bibitem[\protect\citeauthoryear{Meng and Rubin}{1993}]{mengrubin}
\textsc{Meng, X.} and \textsc{Rubin, D.} (1993).
Maximum likelihood estimation via the ECM algorithm: A general framework.
\textit{Biometrika} \textbf{80} 267--278.
\MR{1243503}

\bibitem[\protect\citeauthoryear{Nicolae}{2006a}]{nicolaemd}
\textsc{Nicolae, D.} (2006a).
Quantifying the amount of missing information in genetic association studies.
\textit{Genetic Epidemiology} \textbf{30} 703--717.

\bibitem[\protect\citeauthoryear{Nicolae}{2006b}]{nicolae}
\textsc{Nicolae, D.} (2006b).
Testing untyped alleles (tuna)-applications to genome-wide association studies.
\textit{Genetic Epidemiology} \textbf{30} 718--727.

\bibitem[\protect\citeauthoryear{Purcell et~al.}{2007}]{plink}
\textsc{Purcell, S., Neale, B., Todd-Brown, K., Thomas, L., Ferreira, M., Bender, D.,
  Maller, J., Sklar, P., de~Bakker, P., Daly, M.} and \textsc{Sham, P.} (2007).
Plink: A toolset for whole-genome association and population-based linkage analysis.
\textit{American Journal of Human Genetics} \textbf{81} 559--575.

\bibitem[\protect\citeauthoryear{Sankararaman et~al.}{2009}]{sankararamanetal}
\textsc{Sankararaman, S., Obozinski, G., Jordan, M.} and \textsc{Halperin, E.} (2009).
Genomic privacy and limits of individual detection in a pool.
\textit{Nature Genetics} \textbf{41} 965--967. Epub.

\bibitem[\protect\citeauthoryear{Scheet and Stephens}{2005}]{scheetstephens}
\textsc{Scheet, P.} and \textsc{Stephens, M.} (2005).
A fast and flexible statistical model for large-scale population genotype data: Applications to inferring missing genotypes and haplotype phase.
\textit{American Journal of Human Genetics} \textbf{78} 629--644.

\bibitem[\protect\citeauthoryear{Servin and Stephens}{2008}]{servinstephens}
\textsc{Servin, B.} and \textsc{Stephens, M.} (2008).
Imputation-based analysis of association studies: Candidate regions and quantitative traits.
\textit{PLoS Genetics} \textbf{3} e114.

\bibitem[\protect\citeauthoryear{Stephens and Scheet}{2005}]{stephensscheet}
\textsc{Stephens, M.} and \textsc{Scheet, P.} (2005).
Accounting for decay of linkage disequilibrium in haplotype inference and missing-data imputation.
\textit{American Journal of Human Genetics} \textbf{76} 449--462.

\bibitem[\protect\citeauthoryear{Stephens, Smith and Donnelly}{2001}]{stephenssmithdonnelly}
\textsc{Stephens, M., Smith, N.} and \textsc{Donnelly, P.} (2001).
A new statistical method for haplotype reconstruction from population data.
\textit{American Journal of Human Genetics} \textbf{68} 978--989.

\bibitem[\protect\citeauthoryear{The International HapMap Consortium}{2005}]{hapmap}
\textsc{The International HapMap Consortium} (2005).
A haplotype map of the human genome.
\textit{Nature} \textbf{437} 1299--1320.

\bibitem[\protect\citeauthoryear{Weir}{1979}]{weir79}
\textsc{Weir, B.} (1979).
Inferences about linkage disequilibrium.
\textit{Biometrics} \textbf{35} 235--254.

\bibitem[\protect\citeauthoryear{Wellcome Trust Case Control Consortium}{2007}]{wellcometrust}
\textsc{Wellcome Trust Case Control Consortium} (2007).
Genome-wide association study of 14,000 cases of seven common diseases and 3,000 shared controls.
\textit{Nature} \textbf{447} 661--678.

\bibitem[\protect\citeauthoryear{West and Harrison}{1997}]{westharrison}
\textsc{West, M.} and \textsc{Harrison, J.} (1997).
\textit{Bayesian Forecasting and Dynamic Models}, 2nd ed. Springer, New York.
\MR{1482232}

\bibitem[\protect\citeauthoryear{Yu and Schaid}{2007}]{yuschaid}
\textsc{Yu, Z.} and \textsc{Schaid, D.} (2007).
Methods to impute missing genotypes for population data.
\textit{Human Genetics} \textbf{122} 495--504.

\bibitem[\protect\citeauthoryear{Zeggini et~al.}{2008}]{zegginietal}
\textsc{Zeggini, E., Scott, L., Saxena, R., Voight, B., Marchini, J., Hu, T.,
  de Bakker, P., Abecasis, G., Almgren, P., Andersen, G., Ardlie, K., Bostr\"{o}m,
  K., Bergman, R., Bonnycastle, L., Borch-Johnsen, K., Burtt, N., Chen, H.,
  Chines, P., Daly, M., Deodhar, P., Ding, C., Doney, A., Duren, W., Elliott,
  K., Erdos, M., Frayling, T., Freathy, R., Gianniny, L., Grallert, H., Grarup,
  N., Groves, C., Guiducci, C., Hansen, T., Herder, C., Hitman, G., Hughes, T.,
  Isomaa, B., Jackson, A., J\o rgensen, T., Kong, A., Kubalanza, K., Kuruvilla,
  F., Kuusisto, J., Langenberg, C., Lango, H., Lauritzen, T., Li, Y., Lindgren,
  C., Lyssenko, V., Marvelle, A., Meisinger, C., Midthjell, K., Mohlke, K.,
  Morken, M., Morris, A., Narisu, N., Nilsson, P., Owen, K., Palmer, C., Payne,
  F., Perry, J., Pettersen, E., Platou, C., Prokopenko, I., Qi, L., Qin, L.,
  Rayner, N., Rees, M., Roix, J., Sandbaek, A., Shields, B., Sj\"{o}gren, M.,
  Steinthorsdottir, V., Stringham, H., Swift, A., Thorleifsson, G.,
  Thorsteinsdottir, U., Timpson, N., Tuomi, T., Tuomilehto, J., Walker, M.,
  Watanabe, R., Weedon, M., CJ, C. W., {W}ellcome Trust Case
  Control Consortium, Illig, T., Hveem, K., Hu, F., Laakso, M., Stefansson, K.,
  Pedersen, O., Wareham, N., Barroso, I., Hattersley, A., Collins, F., Groop,
  L., McCarthy, M., Boehnke, M.} and \textsc{Altshuler, D.} (2008).
Meta-analysis of genome-wide association data and large-scale replication identifies additional susceptibility loci for type 2 diabetes.
\textit{Nature Genetics} \textbf{40} 638--645.

\end{thebibliography}
\end{document}